\documentclass{ws-ijmpa}

\usepackage{graphicx}

\begin{document}

\markboth{Carl H. Albright}
{SO(10) GUT Models and Their Present Success in Explaining Mass and Mixing 
Data}

\catchline{}{}{}

\title{SO(10) GUT Models and Their Present Success\\
	in Explaining Mass and Mixing Data\footnote{FERMILAB-Conf-02/326-T,
	paper presented at the Neutrinos and Implications for Physics 
	Beyond the Standard Model Conference, SUNY at Stony Brook, 
	October 11-13, 2002.}}	

\author{\footnotesize CARL H. ALBRIGHT}

\address{Department of Physics, Northern Illinois University\\
	DeKalb, Illinois 60115, USA\\
	and \\
	Fermi National Accelerator Laboratory, P.O. Box 500\\
	Batavia, Illinois 60510, USA\footnote{present address}}

\maketitle

\pub{Received (Day Month Year)}{Revised (Day Month Year)}

\begin{abstract}
Some features of SO(10) GUT models are reviewed, and a number of such 
models in the literature are compared.  While some have been eliminated
by recent neutrino data, others are presently successful in explaining
the quark and lepton mass and mixing data.  A short description of one
very predictive model is given which illustrates some of the features
discussed.  Future tests of the models are pointed out including one 
which contrasts sharply with those models based on an $L_e - L_{\mu} 
- L_{\tau}$ type symmetry.   

\end{abstract}

\section{Introduction}	

Many mass matrix models in the literature\cite{b-d} attempt to explain only 
the recent mass and mixing data in the lepton sector.  More ambitious attempts
introduce supersymmetric grand unified (SUSY GUT) models to understand both 
the lepton and quark sectors.  In this brief review my attention is 
restricted to four-dimensional three family SO(10) GUT models with no light
sterile neutrinos.  One finds that several models are presently quite 
successful in explaining the data, including the preferred LMA solar 
neutrino solution.  One model is illustrated in some detail, while 
future critical tests of the presently successful models are described.

\section{SO(10) Model Structure}

It is well known that the three families of left-handed quarks and leptons
and their left-handed charge conjugates fit neatly into three copies of
the SO(10) spinor representation, ${\bf 16}_i,\ i=1,2,3$.  In 
fact, this feature is what has made SO(10) so attractive as a unification group.
Higgs fields appearing in the ${\bf 45}_H,\ {\bf 16}_H$ and $\overline{\bf 
16}_H$ are needed to break SO(10) to the standard model.  The two 
light Higgs doublets which are required to break the electroweak symmetry 
can be accommodated by a single ${\bf 10}_H$ of SO(10), which 
consists of a $5 + \bar{5}$ of SU(5) or a $(6,1,1) + (1,2,2)$ of SU(4) 
$\times$ SU(2)$_L$ $\times$ SU(2)$_R$. Doublet-triplet splitting of the Higgs 
fields is required and can be achieved 
via the Dimopoulos-Wilczek mechanism,\cite{d-w} if the $\langle {\bf 45}_H 
\rangle$ VEV points in the $B - L$ direction.  With only one ${\bf 10}_H$ 
effecting the electroweak breaking, $\tan \beta \equiv v_u/v_d \sim 55$.  

The above represents the essential ingredients of an SO(10) model.  However,
many authors have found it desirable to introduce additional Higgs fields.
For example, an additional ${\bf 16}'_H,\ \overline{\bf 16}'_H$ pair can
help to stabilize the double-triplet splitting solution.\cite{b-r}  If the VEV
$\langle \bar{5}({\bf 16}'_H) \rangle \neq 0$, the light Higgs doublet, 
$H_d$, can reside in both the vector and spinor representations, i.e.,  
\begin{equation}
	H_d \sim \bar{5}({\bf 10}_H)\cos \gamma + \bar{5}({\bf 16}'_H)
		\sin \gamma,
\label{hdcomb}
\end{equation}
so that Yukawa coupling unification is possible for any value in the range
$\tan \beta \sim 1 - 55$.  Similar results are also possible with the 
addition of a ${\bf 126}_H,\ \overline{\bf 126}_H$ pair in place of the
${\bf 16}'_H,\ \overline{\bf 16}'_H$ pair of Higgs fields.

It may also be desirable to introduce ${\bf 16},\ \overline{\bf 16}$ pairs,
${\bf 10}$'s, etc. of matter fields provided they get supermassive near the
GUT scale.  They can then be integrated out in Froggatt-Nielsen type
diagrams\cite{f-n} which provide higher-order effective interaction 
contributions to the mass matrix elements.

\section{Horizontal Flavor Symmetries}

While SO(10) relates quarks and leptons of one family, it is necessary to 
invoke some horizontal flavor symmetry in order to avoid the bad SU(5) 
relations such as $m_d = m_e$ and $m_s = m_\mu$.  This can be done at four
different levels of model building with the following prescriptions.

\begin{itemlist}
 \item Level 1: Simply impose a certain texture, such as a modified Fritzsch
	form for the mass matrices.
 \item Level 2: Introduce an effective $\lambda \sim 0.22$ expansion for 
	each mass matrix.  The prefactors of the expansion parameters 
	typically are not precisely determined, however.
 \item Level 3: Assign effective operators for each matrix element, possibly
	with some flavor symmetry imposed.
 \item Level 4: Introduce a horizontal flavor symmetry which assigns flavor
	charges to every Higgs and matter superfield.  Higgs and Yukawa
	superpotentials are constructed in terms of renormalizable (and 
	possibly some non-renormalizable) terms which obey that flavor
	symmetry.  Matrix elements then follow from the Froggatt-Nielsen
	diagrams which can be constructed.
\end{itemlist}

\section{Some General Observations} 

\begin{itemlist}
 \item  The SO(10) models found in the literature differ by their choice of 
	Higgs structure, horizontal flavor symmetry and the flavor charge 
	assignments, if any.  
 \item  The desirable Georgi-Jarlskog relations,\cite{g-j} 
	$m_s = m_\mu/3,\ m_d = 3m_e$, 
	can be readily obtained if a $\langle {\bf 45}_H \rangle$ Higgs VEV
	points in the $B - L$ direction, or if a $\langle \bar{5}({\bf 126}_H)
	\rangle$ VEV is involved.  
 \item  The presence of a $\langle \bar{5}({\bf 16}_H)\rangle$ VEV and a 
	flavor symmetry will typically lead to lopsided\cite{lopsided} down 
	quark and charged lepton mass matrices, $D$ and $L$.  This is useful to 
	explain the small $V_{cb}$ and large $U_{\mu 3}$ mixing matrix 
	elements.\cite{lopmix}  A consequence of this lopsided nature is  
	an enhanced flavor-violating $\tau \rightarrow \mu \gamma$ decay rate 
	that is within one or two orders of magnitude of the present 
	experimental limit.  Hence future improved experiments will be able 
	to confirm or rule out this mechanism.
 \item  Most early models were easily able to accommodate the SMA solar 
	neutrino solution, while some could accommodate the LOW or QVO 
	solution as well.  However, to obtain the LMA solution in the SO(10) 
	GUT model framework with the seesaw mechanism, some fine tuning is 
	generally required.  Typically, models which require special features
	of the Dirac and right-handed Majorana mass matrices, $N$ and $M_R$,
	to get maximal atmospheric mixing have trouble getting the LMA solar
	solution.  That is easier to achieve if the $M_R$ matrix can be 
	independently adjusted to yield the LMA solution, while $N$ and $L$
	conspire to give maximal atmospheric mixing.
\end{itemlist}

\section{Some Selected SO(10) Models}

A number of SO(10) SUSY GUT models can be found in the literature.\cite{b-d} 
To illustrate the success of some well-known models, I have confined my 
attention to four-dimensional models with three quark and lepton 
families for which the seesaw mechanism applies with three right-handed 
(conjugate left-handed) singlet neutrino fields.  I have also assumed that the 
presently-preferred LMA solution\cite{solar} will be confirmed by 
KamLAND.\cite{kamland} 
On this basis already some of the models, as constructed, have been ruled out 
by more recent mass and mixing data, while others still survive.  It is 
instructive, however, to compare the various features of all the models
considered.

\begin{table}[h]
\tbl{Features of some selected SO(10) models.}
{\begin{tabular}{@{}lrcllcccc@{}} \toprule
  Model & Ref. & Level & Flavor Sym. & Texture & $\tan \beta$ & CKM & 
	Solar & Viable \\ \colrule
AB & \cite{ab} & 4 & $U(1)\times Z_2\times Z_2$ & Lopsided & $\sim 5$ & Yes & 
	LMA & Yes \\[0.1in]
BPW & \cite{bpw} & 3 & effective & Sym/Asym & low & Yes & LMA & Yes \\
 & & & operators & & & & &\\[0.1in]
BR & \cite{br} & 4 & $SU(3)$ & Lopsided & 1-10 & Yes & SMA & No \\[0.1in]
BRT & \cite{brt} & 4 & $U(2)\times U(1)^n$ & Sym/Asym & $\sim 55$ & No & LMA & 
	No \\[0.1in]
BW & \cite{bw} & 1 & postulated & Sym & ? & Yes & LMA & ? \\[0.1in]
CM & \cite{cm} & 4 & $U(2)\times (Z_2)^3$ & Sym & 10 & Yes & LOW & No \\[0.1in]
CW & \cite{cw} & 4 & $\Delta(48)\times U(1)$ & Sym/Asym & $\sim 2$ & Yes & LMA 
	& No \\[0.1in]
KM & \cite{km} & 2 & $SU(3)\times U(1)$ & Lopsided & small & ? & LMA & ? 
	\\[0.1in]
M & \cite{m} & 2 & $U(1)_A \times Z_2$ & Lopsided & small & Yes & LMA & No 
	\\[0.1in]
RV-S & \cite{rvs} & 2 & $SU(3)$ and & Sym/Asym & ? & Yes & LMA & Yes \\
 & & & Abelian & \\ \botrule
\end{tabular}}
\end{table}

Table 1	lists the models with their level of construction, flavor symmetry, 
texture, applicable range of $\tan \beta$, whether or not they fit the 
CKM mixing matrix and their preferred solar neutrino solution.  Some textures 
correspond to lopsided 
mass matrices, while others have only symmetric or both symmetric and 
antisymmetric entries.  The latter favor large values of $\tan \beta$ to give 
the desired Yukawa coupling unification at the GUT scale, while the lopsided 
models tend to require low or moderate values of $\tan \beta$ in order that the 
matrices be lopsided enough.  Thus the determination of $\tan \beta$, as 
well as the observation of the $\tau \rightarrow \mu \gamma$ mentioned 
earlier, will serve to rule out one choice or the other.  

Some features of the models warrant specific remarks.  
In the Blazek-Raby-Tobe model,\cite{brt} a sterile neutrino is present while 
the apex of the CKM triangle is 
in the second quadrant which is now at odds with the recent quark mixing 
determination.  The Chou-Wu model\cite{cw} requires a sterile neutrino
to get the solar LMA solution.  The Chen-Mahanthappa model\cite{cm} prefers the
LOW solar solution, for the LMA solution can not be obtained without violating 
the upper CHOOZ bound\cite{CHOOZ} on $U_{e3}$.  The Maekawa model\cite{m} also 
violates the CHOOZ bound on $U_{e3}$.  For the Buchmuller-Wyler model\cite{bw}
and the Kitano-Mimura model\cite{km}, it is not clear from their solutions 
whether the LMA mixing is in the presently allowed range.  Of the three 
remaining apparently successful models, the Babu-Pati-Wilczek model\cite{bpw} 
requires a non-seesaw 
contribution to the left-handed Majorana matrix, $M_L$, in order to fit both 
the atmospheric and solar LMA solutions.  The Ross-Velasco-Sevilla 
model\cite{rvs}
is rather recent and has not been completely specified.  To illustrate
some of the features of SO(10) models cited earlier, some detailed features
of the very predictive Albright-Barr model\cite{ab} are presented in the next 
Section.

Of the models listed in Table 1, some are already essentially ruled out 
by the more accurate recent quark and lepton mixing data, but as we have seen,
several are still viable.  In making this judgment I have assumed there are 
no light sterile neutrinos and that the LMA solution is the 
correct one.  Of course, some models which are on the verge of being ruled
out may be revived by their authors with further adjustments.

\section{Example of the LMA Solution in One Predictive SO(10) Model}

The model developed in Ref. 10 is based on a $U(1)\times Z_2 \times Z_2$
flavor symmetry that stabilizes the Dimopoulos-Wilczek solution to the 
doublet-triplet splitting problem by the introduction of a second pair 
of ${\bf 16}_H,\ \overline{16}_H$ Higgs fields.\cite{b-r}  The Higgs
and Yukawa superpotentials can be written down after flavor charges for that 
symmetry are assigned to all the Higgs and matter fields.  
The mass matrices then follow from Froggatt-Nielsen diagrams involving 
the vertex terms appearing in the superpotentials.  

The Dirac mass matrices for the up and down quarks, neutrinos and charged
leptons are found to be  

\begin{equation}
$$\begin{array}{ll}
\hspace*{-0.6in} U = \left( \begin{array}{ccc} \eta & 0 & 0 \\
  0 & 0 & \epsilon/3 \\ 0 & - \epsilon/3 & 1 \end{array} \right)M_U,\ 
  & D = \left( \begin{array}{ccc} \eta & \delta & \delta' e^{i\phi}
  \\
  \delta & 0 & \sigma + \epsilon/3  \\
  \delta' e^{i \phi} & - \epsilon/3 & 1 \end{array} \right)M_D, \\ & \\
\hspace*{-0.6in} N = \left( \begin{array}{ccc} \eta & 0 & 0 \\ 0 & 0 & 
  - \epsilon \\ 0 & \epsilon & 1 \end{array} \right)M_U,
  & L = \left( \begin{array}{ccc} \eta & \delta & \delta' e^{i \phi} \\
  \delta & 0 & -\epsilon \\ \delta' e^{i\phi} & 
  \sigma + \epsilon & 1 \end{array} \right)M_D,\\[0.4in]
\end{array}$$
\label{matrices}
\end{equation}

\noindent
Several texture zeros appear in elements for which the flavor symmetry
forbids the appearance of any Froggatt-Nielsen diagrams.  The antisymmetric
$\epsilon$ terms arise from diagrams involving the adjoint 
$\langle {\bf 45}_H \rangle$ Higgs VEV pointing in the $B - L$ direction.  
The lopsided nature of the large $\sigma$ terms in $D$ and $L$ arises from 
the appearances of diagrams involving the $\langle \bar{5}({\bf 16}_H) \rangle$
Higgs VEV as suggested earlier.  

The eight input parameters are defined at the GUT scale and are set equal to

\begin{equation}
$$\begin{array}{rlrl}
        M_U&\simeq 113\ {\rm GeV},&\qquad M_D&\simeq 1\ {\rm GeV},\\
        \sigma&=1.78,&\qquad \epsilon&=0.145,\\
        \delta&=0.0086,&\qquad \delta'&= 0.0079,\\
        \phi&= 54^o,&\qquad \eta&= 8 \times 10^{-6}.\\
\end{array}$$
\label{input}
\end{equation}

\noindent With these values, the structures of the $D$ and $L$ matrices
lead to the Georgi-Jarlskog relations at the GUT scale with Yukawa coupling 
unification for $\tan \beta \sim 5$.  All nine quark and charged lepton 
masses plus the three CKM angles and CP phase are well-fitted with these 
input parameters after evolution from the GUT scale:

\begin{equation}
$$\begin{array}{rlrl}
           m_t(m_t) &= 165\ {\rm GeV},\quad & m_{\tau} &= 1.777\ {\rm GeV}
                \\[0.1in]
           m_u(1\ {\rm GeV}) &= 4.5\ {\rm MeV},\quad & m_\mu &= 105.7\ 
                {\rm MeV}\\[0.1in]
           V_{us} &= 0.220, \quad & m_e &= 0.511\ {\rm MeV}\\[0.1in]
           V_{cb} &= 0.0395, \quad & \delta_{CP} &= 64^\circ\\
          \end{array}$$
\label{inputvalues}
\end{equation}

\noindent determine the input parameters which lead to the following 
predictions:

\begin{equation}
$$\begin{array}{rlrl}
           m_b(m_b) &= 4.25\ {\rm GeV},\quad & m_c(m_c) &= 1.23\ {\rm GeV}
                \\[0.1in]
           m_s(1\ {\rm GeV}) &= 148\ {\rm MeV},\quad & m_d(1\ {\rm MeV}) 
                &= 7.9\ {\rm MeV}\\[0.1in]
           |V_{ub}/V_{cb}| &= 0.080,\quad & \sin 2\beta &= 0.64.\\
          \end{array}$$
\label{predictions}
\end{equation}

\noindent The Hermitian matrices $U^\dagger U,\ D^\dagger D$, and 
$N^\dagger N$ are diagonalized by small LH rotations, while 
$L^\dagger L$ is diagonalized by a large LH rotation.  This accounts for
the fact that $V_{cb} = (U^\dagger_U U_D)_{cb}$ is small,
while $U_{\mu 3} = (U^\dagger_L U_\nu)_{\mu 3}$ is large and responsible for 
the maximal atmospheric neutrino mixing for any reasonable $M_R$.

The type of $\nu_e \leftrightarrow \nu_\mu,\nu_\tau$ solar neutrino mixing 
is determined by the texture of $M_R$, since the solar and 
atmospheric mixings are essentially decoupled in this model.
Further study reveals the LMA solution requires a nearly hierarchical 
texture\cite{ab} which can also be understood with Froggatt-Nielsen 
diagrams.  The texture suggested is 

\begin{equation} 
	M_R = \left(\matrix{b^2 \eta^2 & -b\epsilon\eta & a\eta\cr 
		-b\epsilon\eta & \epsilon^2 & -\epsilon\cr
		a\eta & -\epsilon & 1\cr}\right)\Lambda_R,
\label{MR}
\end{equation}

\noindent with the parameters $\epsilon$ and $\eta$ specified in Eq. 
(\ref{input}).  Here $\Lambda_R$ then sets the scale of the heavy right-handed 
Majorana neutrino masses and determines $\Delta m^2_{32}$ for the atmospheric 
neutrino mixing by the seesaw mechanism.  

\begin{figure*}[b]
\centering\leavevmode
\includegraphics*[width=2.4in]{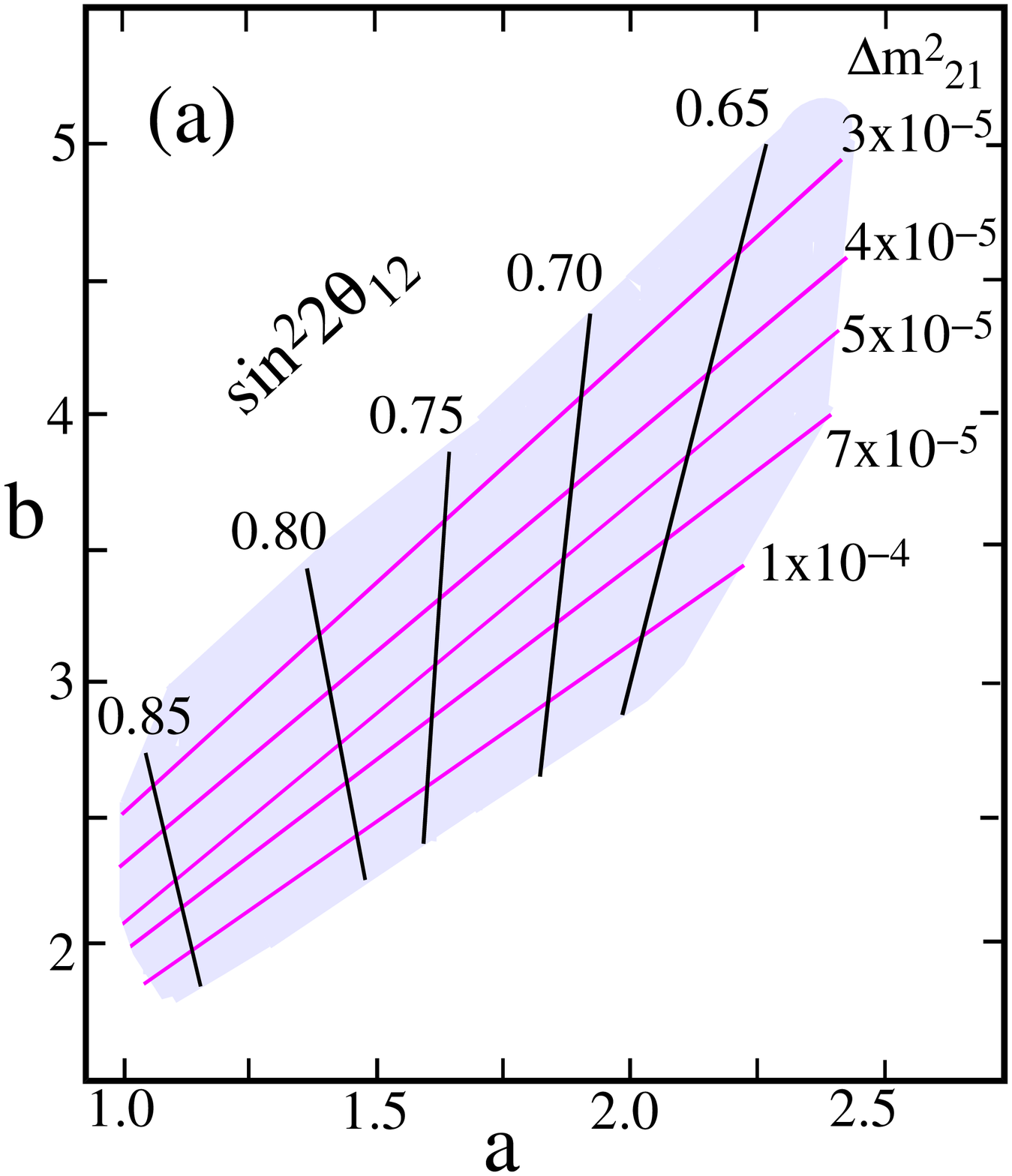}
\hspace*{0.0in}
\includegraphics*[width=2.4in]{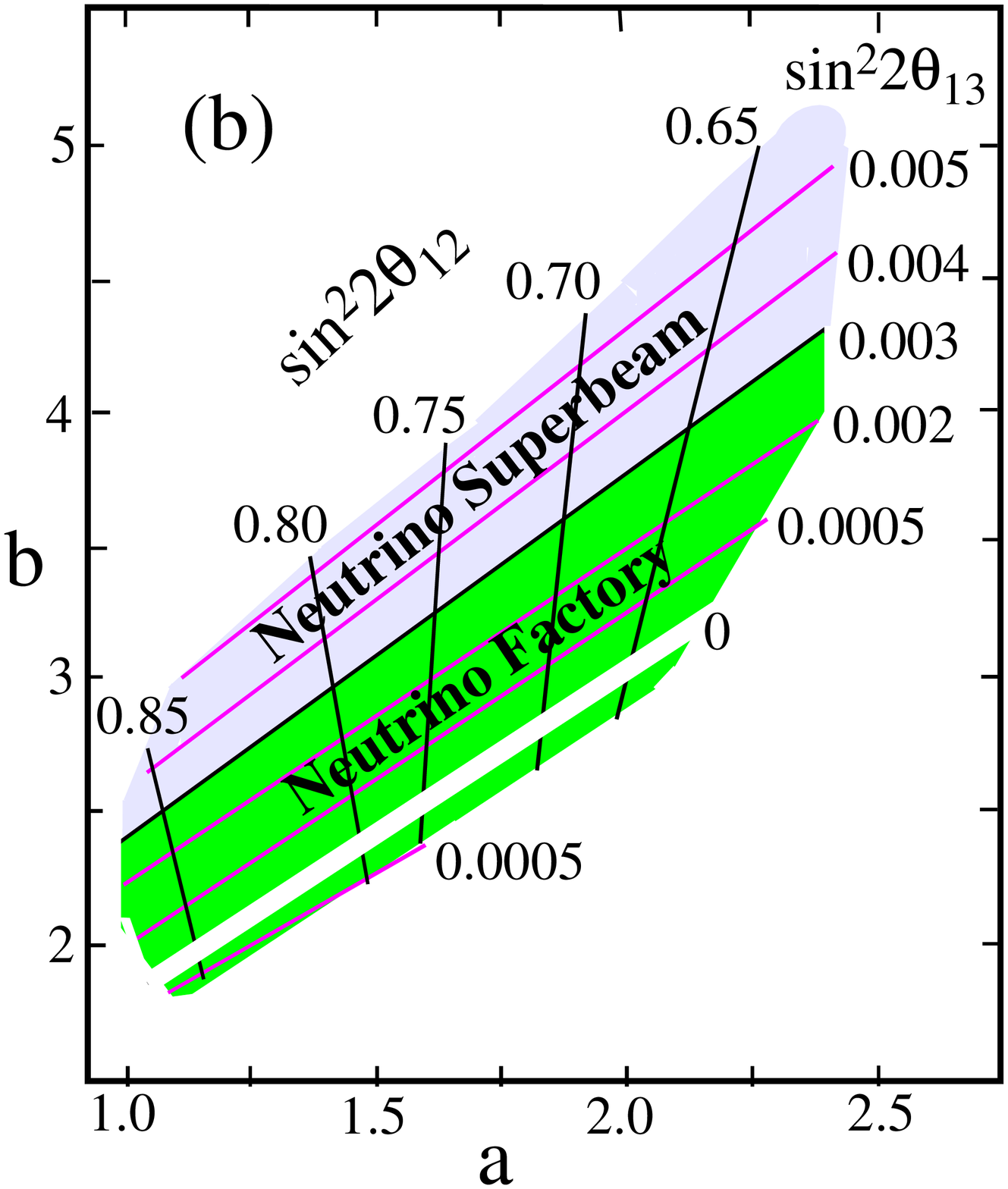}
\vspace*{-0.1in}
\caption[]{The viable region of GUT parameter space consistent 
with the present bounds on the LMA MSW solution.  Contours of 
constant $\sin^2 2\theta_{12}$ are shown together with (a) contours of 
constant $\Delta m^2_{21}$ and (b) contours of $\sin^2 2\theta_{13}$.}
\label{fig:gut_plot}
\end{figure*}

The allowed parameter space in the $a - b$ plane\cite{ag} shown in Fig. 1 
corresponds to the pre-SNO allowed LMA solar neutrino region\cite{s-k} in the 
$\Delta m^2_{21} - \sin^2 2\theta_{12}$ mixing plane.  With the recent SNO and 
Super-Kamiokande results,\cite{solar} it should be understood that part of the 
allowed parameter region corresponding to higher values of $a$, i.e., lower 
values of $\sin^2 2\theta_{12}$, has been eliminated.  In Fig. 1(a)
contours of constant $\sin^2 2\theta_{12}$ and $\Delta m^2_{21}$ are shown,
while contours of constant $\sin^2 2\theta_{12}$ and $\sin^2 2\theta_{13}$
are given in Fig. 1(b).  Once $\Delta m^2_{21}$ and $\sin^2 2\theta_{12}$ are
known, the model parameters $a$ and $b$ are determined by Fig. 1(a) from 
which the reactor neutrino mixing $\sin^2 2\theta_{13}$ can be found
from Fig. 1(b).  We observe that the reactor angle, $\theta_{13}$, as determined
in this model is generally much smaller than that determined from the 
present CHOOZ bound,\cite{CHOOZ} i.e., $|U_{e3}| \simeq \sin \theta_{13} 
< 0.16$ or $\sin^2 2\theta_{13} < 0.10$.  As indicated, a Neutrino Factory 
will be required to determine $\theta_{13}$ for a large part of the 
presently allowed region.

As an interesting special case, we note that with $a = 1,\ b = 2\ {\rm and}\ 
\Lambda_R = 2.72 \times 10^{14}$ GeV, the seesaw mechanism leads to the
simple light neutrino mass matrix

\begin{equation}
	M_\nu = \left(\matrix{ 0 & -\epsilon & 0\cr 
			-\epsilon & 0 & 2\epsilon\cr 0 & 2\epsilon & 1\cr}
			\right)M^2_U/\Lambda_R.
\label{spmatrix}
\end{equation}

\noindent From $M_\nu$, $L$ and the input parameters we then find

\begin{equation}
$$\begin{array}{l}
          M_1 = 3.2 \times 10^{8},\ M_2 = 3.6 \times 10^8,\  M_3 = 2.8 
		\times 10^{14}\ {\rm GeV},\\[0.1in]
	  m_1 = 4.9\ {\rm mev},\quad m_2 = 8.7\ {\rm mev},
		\quad m_3 = 51\ {\rm mev},\\[0.1in]
          \Delta m^2_{32} = 2.5 \times 10^{-3}\ {\rm eV^2},\quad 
		\sin^2 2\theta_{\rm atm} = 0.994,\\[0.1in]
          \Delta m^2_{21} = 5.1 \times 10^{-5}\ {\rm eV^2},\ \sin^2 
		2\theta_{\rm sol} = 0.88,\ \tan^2 \theta_{13} = 0.49,\\[0.1in]
          U_{e3} = -0.014,\quad \sin^2 2\theta_{\rm reac} = 0.0008\\
	  \end{array}$$
\label{nupred1}
\end{equation}

\noindent Note that $\Lambda_R$ has not only set the scale for the atmospheric
neutrino mixing $\Delta m^2_{32}$ but also for the solar neutrino mixing 
$\Delta m^2_{21}$.  For example, a value of $\Delta m^2_{32} = 2.8 \times 
10^{-3}\ {\rm eV}^2$ results in $\Delta m^2_{21} = 5.7 \times 10^{-5}\ {\rm
eV}^2$, while the mixing angles remain unchanged.  Whereas one might have 
expected an inverted hierarchy 
with $M_1$ and $M_2$ so close together and much smaller than $M_3$, the 
resultant form of $M_\nu$ leads to a normal but rather mild hierarchy for 
the light left-handed neutrino masses.

\section{Future Tests of SO(10) Models}

Several critical tests will be made in the future with long baseline 
experiments involving Superbeams, and possibly Neutrino Factories.  These 
tests involve the nature of the light neutrino
mass hierarchy, i.e., normal vs. inverted; the value of the reactor neutrino
mixing angle $\theta_{13}$ or the element $|U_{e3}| \simeq \sin \theta_{13}$;
and the determination of the leptonic Dirac CP-violating phase $\delta$.  
For the three models considered which clearly appear to be still viable, the 
predictions are listed in Table 2.

\begin{table}[t]
\tbl{Predictions of the presently successful models considered.}
{\begin{tabular}{@{}lccccc@{}} \toprule
  Model & Ref. & Hierarchy & $|U_{e3}|$ & $\sin^2 2\theta_{13}$ & CP Violation\\
	\colrule
AB & \cite{ab} & Normal & 0 - 0.035 & 0 - 0.005 & Small \\[0.1in]
BPW & \cite{bpw} & Normal & ? & ? & ?\\[0.1in]
RV-S & \cite{rvs} & Normal & $\sim 0.07$ & $\sim 0.02$ & ? \\[0.1in]
\botrule
\end{tabular}}
\end{table}

It is apparent that the presently successful SO(10) GUT models favor a 
normal hierarchy.  This is in stark contrast with the models with 
a conserved lepton number quantity,\cite{zee} such as $L_e - L_{\mu} - 
L_{\tau}$, which favor an inverted hierarchy.\footnote{For a recent variant of 
such models and additional references, see Ref. 24.}  On the other hand, the 
predicted value for $|U_{e3}|$ is apparently quite model dependent, 
with some models predicting values very close to the CHOOZ bound,
while others require a Neutrino Factory to pin down the correct
value.  Unfortunately, the leptonic CP violating phase $\delta$, which is of 
great interest if the LMA solution is the correct one, is not well determined 
in most models. 
 
\section{Summary}

A number of $SO(10)$ SUSY GUT models have been proposed in the literature
with a small but interesting sample considered here.  Some have been, or are 
on the verge of being, eliminated, while others still survive and are
able to explain all the known quark and lepton mass and mixing data. 
Long baseline experiments which can determine whether the neutrino mass 
hierarchy is normal or inverted appear to have a direct bearing on the 
survival of SO(10) vs. nearly-conserved $L_e - L_\mu -L_\tau$ type models.  
This particular test appears to be one of the most promising for narrowing 
down the list of successful model candidates.  

The observed value of $\sin^2 2\theta_{13}$ appears to be less discriminatory
between models of the SO(10) or the conserved lepton type.  Some models of 
both types predict that $\theta_{13}$ lies just below the CHOOZ bound and will 
be observable with off-axis beams and/or Superbeams.  Others favor such low 
values of $\theta_{13}$ that a Neutrino Factory will be required to determine 
its value.
  
The issue of proton decay via dim-5 operators is potentially a serious one
for GUT models, if proton decay is not detected shortly.\cite{raby}  On the 
other hand, by formulating an SO(10) model in five dimensions, one can 
eliminate the dim-5 operator contributions entirely.\cite{ab5}  The dim-6 
operators will still be present and possibly somewhat enhanced, but they 
typically lead to lifetimes for proton decay which are presently two to 
three orders of magnitude larger than the present lower bounds.

\section*{Acknowledgements}

The author thanks the Fermilab Theoretical Physics Department for its kind
hospitality while this work was work was carried out.  Fermilab is operated
by Universities Research Association, Inc. under Contract No. DE-AC02-76CH03000 
with the Department of Energy.


\begin{thebibliography}{0}

\bibitem{b-d}	S.M. Barr and I. Dorsner, {\it Nucl. Phys.} {\bf B585}, 79
		(2000); G. Altarelli and F. Feruglio, hep-ph/0206077;  
		S.F. King, hep-ph/0208266. 

\bibitem{d-w}	S. Dimopoulos and F. Wilczek, Report No. NSF-ITP-82-07, 1981,
		in {\it Proceedings of the 19th Course of the International
		School of Subnuclear Physics, Erice, Italy, 1981}, ed. 
		A. Zichichi (Plenum Press, New York, 1983).

\bibitem{b-r}	S.M. Barr and S. Raby, {\it Phys. Rev. Lett.} {\bf 79}, 4748
		(1997).

\bibitem{f-n}	C.D. Froggatt and H.B. Nielsen, {\it Nucl. Phys.} {\bf B147},
		277 (1979). 

\bibitem{g-j}	H. Georgi and C. Jarlskog, {\it Phys. Lett.} {\bf B86}, 297
		(1979).

\bibitem{lopsided} K.S. Babu and S.M. Barr, {\it Phys. Lett.} {\bf B381},
		202 (1996). 

\bibitem{lopmix} J. Sato and T. Yanagida, {\it Phys. Lett.} {\bf B430}, 
		127 (1998); C.H. Albright, K.S. Babu and S.M. Barr, 
		{\it Phys. Rev. Lett.} {\bf 81}, 1167 (1998);  N. Irges,
		S. Lavignac and P. Ramond, {\it Phys. Rev.} {\bf D58}, 
		035003 (1998).

\bibitem{solar}	Super-Kamiokande Collab., {\it Phys. Lett.} {\bf B539}, 
		179 (2002); SNO Collab., {\it Phys. Rev. Lett.} {\bf 89},
		011302 (2002). 

\bibitem{kamland} S.A. Dazeley for KamLAND Collab., hep-ex/0205041. 

\bibitem{ab}	C.H. Albright and S.M. Barr, {\it Phys. Rev. Lett.} {\bf 85},
		244 (200); {\it Phys. Rev.} {\bf D62}, 093008 (2000); 
		{\it Phys. Rev.} {\bf D64}, 073010 (2001). 

\bibitem{bpw}	K.S. Babu, J.C. Pati and F. Wilczek, {\it Nucl. Phys.} 
		{\bf B566}, 33 (2000); J.C. Pati, hep-ph/0209160. 

\bibitem{br}	Z. Berezhiani and A. Rossi, {\it Nucl. Phys.} {\bf B594},
		113 (2001).

\bibitem{brt}	T. Blazek, S. Raby and K. Tobe, {\it Phys. Rev.} {\bf D62},
		055001 (2000).

\bibitem{bw}	W. Buchmuller and D. Wyler, {\it Phys. Lett.} {\bf B521},
		291 (2001).

\bibitem{cm}	M.-C. Chen and K.T. Mahanthappa, {\it Phys. Rev.} {\bf D65},
		053010 (2002).

\bibitem{cw}	K.C. Chou and Y.L. Wu, in {\it Proceedings of the Symposium on
		Frontiers of Physics at Millenium, Beijing, 1999}, ed. Y.L. 
		Wu and J.P. Hsu, (World Scientific, Singapore, 2001).

\bibitem{km}	R. Kitano and Y. Mimura, {\it Phys. Rev.} {\bf D63}, 016008
		(2001).

\bibitem{m}	N. Maekawa, {\it Prog. Theor. Phys.} {\bf 106}, 401 (2001).

\bibitem{rvs}	G.G. Ross and L. Velasco-Sevilla, hep-ph/0208218.

\bibitem{CHOOZ} CHOOZ Collab., {\it Phys. Lett.} {\bf B420}, 397 (1998).

\bibitem{ag}	C.H. Albright and S. Geer, {\it Phys. Rev.} {\bf D65}, 
		073004; {\it Phys. Lett.} {\bf B532}, 311 (2002). 

\bibitem{s-k}	Super-Kamiokande Collab., {\it Phys. Rev. Lett.} {\bf 86},
		5656 (2001).

\bibitem{zee}	S. Petcov, {\it Phys. Lett.} {\bf B110}, 245 (1982).

\bibitem{moh}	R. Kuchimanchi and R.N. Mohapatra, hep-ph/0207373.

\bibitem{raby}	S. Raby, hep-ph/0211024.

\bibitem{ab5}	C.H. Albright and S.M. Barr, hep-ph/0209173.

\end{thebibliography}
\end{document}